\documentclass[11pt]{article}
\usepackage{amssymb, epsfig, amsmath,verbatim,graphicx,mathrsfs}
\usepackage{natbib}
\title{Options to Receive Retirement Gratuity}
\author{Reason L. Machete\footnote{Corresponding author: email: r.l.machete@lse.ac.uk; rmachete@bitri.co.bw}\\\\
{\small Botswana Institute for Technology Research and Innovation}\\
{\small Centre for the Analysis of Time Series, London School of Economics}}
\date{April 17, 2019}
\begin{document}
\maketitle
\noindent\rule{14.6cm}{0.4pt}
\begin{abstract}
Retirement gratuity is the money companies typically pay their employees at the end of their contracts or at the time of leaving the company. It is a defined benefit plan and is often given as an alternative to a pension plan. In Botswana, there is now a new pattern whereby companies give their employees the option to receive their gratuity at various stages before the end of their contracts. In particular, some companies give their employees an option to receive their gratuity on a monthly basis rather than having them wait for a year or more. Many employees find this option attractive, but is it economically sound? This paper sheds light on this question by quantifying the economic benefits of the tax relief provided by government relative to investing the monthly-received funds in a risk-free savings account or helping repay a loan. The principles and methods used herein can be adapted and applied to different taxation systems.
\end{abstract}
\noindent\rule{14.6cm}{0.4pt}
\section{Introduction}
 Retirement gratuity is a defined benefit plan, where retirement time can be thought of as the end of the contract~\citep{jos-11}, but literature on it is rare. It is generally calculated as fraction of an employee's basic multiplied by number of years in service~\citep{yehudah-2002}. In Botswana, some companies offer employment gratuity to their staff when the companies are still new, having no pension fund in place. In other cases, companies offer gratuity-linked contracts as an alternative to being in a pension plan. With uncertainties surrounding pension funds~\citep[e. g. see][]{sta-06}, the idea of receiving one's gratuity at the end of the contract (usually not exceeding five years) can appear attractive. Five years is a relatively short time compared to the long waiting times associated with pension schemes. Having early access to retirement benefits might be consistent with pension systems that favour early retirement even though in such pension systems life long benefits are provided for though not assured~\citep{neu-09}. To governments that are concerned about social security, shortening the time for one to lay hands on the retirement benefits might be objectionable even if politically expedient. The problem with people receiving their retirement benefits early is that they might spend it at the expense of laying something in store for the retirement years.

 In spite of this, some companies in Botswana give employees the option to receive their gratuity monthly, annually, biannually or at the end of their contracts after possibly five years. For certain salary brackets, if the gratuity is received earlier than a year, it is all taxed at a rate of $\delta$. If it is received after at least one year, a proportion $q$ is exempt from tax whilst the rest of the amount is taxed at the rate of $\delta$. Such a use of tax relief is a typical incentive for people to put off some funds towards a retirement plan~\citep[e. g. see][]{fen-14}. The offer to receive one's gratuity monthly in equal installments over one year can, however, have an irresistible appeal as was noted under the superannuation scheme~\citep{fen-14}. \cite{bal-14} noted optimism as a factor that reduces people's probability to put off some funds towards retirement. In Botswana, among the reasons given by some employees for opting to receive their gratuity by monthly installments are the following: (1) To have control over one's finances and determine what to do with it and (2) To save it for oneself and earn interest rather than have it sitting in some unknown account earning no interest to the owner (3) It increases one's monthly salary and thus gives leverage to qualify for loans.

 Whether or not the above reasons are credible is a matter we leave for the reader to decide. Focusing on the cases where one wants to take their gratuity and invest it in a risk-free savings account or to help repay a loan, the issue we seek to address is whether it is economically sound to receive the gratuity monthly or annually. As far as benefiting from tax relief is concerned, there is no real advantage in waiting for more than one year before collecting the gratuity. At the worst, potential interest $r$ that could be gained by placing the accrued gratuity in a bank is lost. In the next section, the problem of how to receive gratuity is considered under simple interest and continuously compounded interest. Some general results for an arbitrary number of equal installments are presented, but the point of this paper is illustrated with specific values of $q$ and $\delta$ to determine the necessary risk-free interest rate $r$. In other countries, there may be several threholds that taxed at different rates. The case considered here is applicable to Botswana and can be analysis can be generalised to multiple threshold scenarios. The final section gives conclusion and implications of the results.
\section{Analysis of the Options}
Suppose the total amount of gratuity that one can accrue in one year is $G$. If one opts to receive the gratuity after exactly one year, the net amount received after tax is $[q+(1-q)(1-\delta)]G$. We will analyse how this compares with being given a lump sum of the gratuity at the beginning of the year and with being given equal installments over one year. We also consider the case where one takes a loan with the hope of using the gratuity as additional funds to help repay the loan.
\subsection{Lump Sum}
To illustrate a point, let us consider the hypothetical case where one receives all their one year's worth of gratuity at the beginning of the year. In this hypothetical case, the net amount of gratuity received after tax is $(1-\delta)G$. One can then take this gratuity and invest it in a risk-free account to earn a simple interest at a rate of $r$ per annum. After one year, the investment account will contain $(1-\delta)(1+r)G$. Bench-marking against a risk-free investment is a concept that is also supported by~\cite{bab-15}, who bench-marked the performance of pension fund cash outflows gainst US Treasury securities.

The crucial question to ask now is: What interest rate should the bank be willing to give if one is to opt to receive the gratuity at the beginning of the year? This question can be addressed by solving the equation
\begin{align}
[q+(1-q)(1-\delta)]G=(1-\delta)(1+r)G.
\label{eqn:gr1}
\end{align}
Making $r$ the subject of the above equation gives the unexpectedly-compact formula
\begin{align}
r=\frac{q\delta}{1-\delta}.
\label{eqn:gr2}
\end{align}
In the particular case of Botswana, $q=1/3$, and for employees earning more than $P12 000$ per month, $\delta=1/4$. Substituting these values into Equation~(\ref{eqn:gr2}) yields $r=1/9$. That is $r\approx 11.11\%$. This result implies that receiving one year's worth of gratuity at the beginning of the year should be attractive provided there is a bank willing to give an interest not less than 11.11\%.

What we just considered is the case when the bank gives simple interest. Alternatively, we can consider the case where the bank gives continuously compounded interest for the whole year, again at a rate of $r$. In this case, Equation~(\ref{eqn:gr1}) is replaced by
\begin{align}
[q+(1-q)(1-\delta)]G=(1-\delta)Ge^r.
\label{eqn:gr3}
\end{align}
This equation yields
\begin{align}
r=\log\left\{[1-(1-q)\delta)]/(1-\delta)\right\}.
\label{eqn:gr4}
\end{align}
Plugging in $q=1/3$ and $\delta=1/4$ into Equation~(\ref{eqn:gr4}) gives $r=\log(10/9)$. That is $r\approx 10.54\%$. This is the minimum interest that should make taking a year's worth of gratuity at the beginning of the year attractive. Of course, one should confirm that such an interest is to be continuously compounded.
\subsection{Equal Installments}
Let us now consider the case in which one receives the gratuity in $n$ equal installments over one year. In this case, at each instant one receives an amount equal to $G/n$ before tax. The actual amount received after tax at every instant is then $(1-\delta)G/n$.
\subsubsection{Simple Interest}
First, we consider the case where the bank offers simple interest at a rate of $r$ per annum. If the gratuity is received at the time instant $t$, where $t\in\{1,\ldots,n\}$, then the corresponding installment becomes
\begin{align}
(1-\delta)\left\{1+\frac{n-t+1}{n}r\right\}G/n
\label{eqn:gr5}
\end{align}
at the end of the year. In particular, the first ($t=1$) installment matures to become $(1-\delta)(1+r)G/n$, the second $(t=2)$ installment matures to $(1-\delta)[1+(n-1)r/n]G/n$, whilst the last ($t=n$) installment matures to $(1-\delta)(1+r/n)G/n$. Adding up all the installments at maturity yields the sum
\begin{align*}
S_n=(1-\delta)(1+r/n)G/n+\ldots+(1-\delta)(1+nr/n)G/n.
\end{align*}
The above formula can equivalently be written as
\begin{align*}
S_n=(1-\delta)G\left[1+\sum_{k=1}^n\frac{kr}{n^2}\right].
\end{align*}
This formula is an arithmetic series and can thus be reduced to the form
\begin{align*}
S_n=(1-\delta)G\left[1+\left(\frac{n+1}{2n}\right)r\right].
\end{align*}
It follows that to find the minimum interest rate that the bank should be willing to give in order to make receiving gratuity by installments attractive, we need to solve the equation
\begin{align*}
[q+(1-q)(1-\delta)]G=(1-\delta)G\left[1+\left(\frac{n+1}{2n}\right)r\right].
\end{align*}
Hence, the minimum interest rate to consider is given by the compact formula
\begin{align}
r=\frac{2nq\delta}{(n+1)(1-\delta)}.
\label{eqn:gr6}
\end{align}
Setting $n=1$ recovers Equation~(\ref{eqn:gr2}), which is when one receives a lump sum at the beginning of the year. In the case where $n=12$, $q=1/3$ and $\delta=1/4$, we get $r=8/39$. That is $r\approx 20.51\%$. Thus, receiving gratuity monthly should be attractive provided the bank is willing to give a simple interest rate of at least $20.51\%$
\subsubsection{Compound Interest}
Consider the case where each of the $n$ installments is invested in an account that gives continuously compounded interest rate $r$. In this case, the installment corresponding to the instant $t$ accrues to
\begin{align}
(1-\delta)\exp{[(n-t+1)r/n]}G/n
\label{eqn:gr7}
\end{align}
at the end of the year. The sum of the installments at maturity is given by the geometric series
\begin{align*}
H_n=(1-\delta)\frac{G}{n}\sum_{k=1}^n\exp{(kr/n)}.
\end{align*}
This series can be compactly written as
\begin{align*}
H_n=(1-\delta)\frac{\exp{(r/n)}\left[\exp{(r)}-1\right]}{[\exp{(r/n)}-1]}\frac{G}{n}.
\end{align*}
The minimum interest rate necessary for one to opt to take gratuity in $n$ equal installments is then found by solving the equation
\begin{align}
[1-(1-q)\delta]G=(1-\delta)\frac{\exp{(r/n)}\left(\exp{(r)}-1\right)}{(\exp{(r/n)}-1)}\frac{G}{n}.
\label{eqn:gra8}
\end{align}
Equation~(\ref{eqn:gra8}) is nonlinear in $r$ and does not have a closed form solution. Nonetheless it can be solved numerically for specific values of $q$, $\delta$ and $n$. Solving it for $n=12$, $q=1/3$ and $\delta=1/4$ gives $r\approx 19.17\%$, which is less than that obtained in the simple interest case by approximately $1.34\%$.
\subsection{Loan Repayment}
Let us now turn to the case where one decides to take a loan amounting to $L$, to be re-paid in $m$ years at an interest of $r_c$ per annum. Assume that in a given year he can repay the amount in $n$ equal installments, the same frequency with which he can receive his gratuity. The interest $r_c$ is compounded at the same frequency of the installments. At this interest rate, the total amount he will have repayed to the bank after m years can be taken to be $R$. Each installment is then $R/(mn)$. At the $k^{\mbox{th}}$ installment, the reduction in the outstanding amount is
\begin{align*}
\left(\frac{R}{mn}-\frac{r_c}{n}L\right)\left(1+\frac{r_c}{n}\right)^{k-1}.
\end{align*}
The sum of the reductions up to the end of the loan period is therefore
\begin{align*}
R_{m,n}=\left(\frac{R}{mn}-\frac{r_c}{n}L\right)\sum_{k=1}^{mn}\left(1+\frac{r_c}{n}\right)^{k-1}.
\end{align*}
Since this is a geometric series, it can be re-written as
\begin{align*}
R_{m,n}=\frac{n}{r_c}\left(\frac{R}{mn}-\frac{r_c}{n}L\right)\left[\left(1+\frac{r_c}{n}\right)^{mn}-1\right].
\end{align*}
At the end of the loan period, the sum of all the reductions will be equal to the loan amount, i.e. $R_{m,n}=L$. Hence we must have
\begin{align*}
L=\frac{n}{r_c}\left(\frac{R}{mn}-\frac{r_c}{n}L\right)\left[\left(1+\frac{r_c}{n}\right)^{mn}-1\right].
\end{align*}
It follows that the total amount repaid at the end of the loan period is
\begin{align}
R=\frac{mr_cL\left(1+\frac{r_c}{n}\right)^{mn}}{\left(1+\frac{r_c}{n}\right)^{mn}-1}.
\label{eqn:gr8}
\end{align}

Suppose now that one wants to use the gratuity to help repay the loan. Considering the first year of the loan, should one wait for a year to use the gratuity after tax relief to repay the loan or is it better to receive the gratuity in $n$ installments to help repay the loan? In each of the two cases, the installments of $R/(mn)$ are committed to the loan. If one decides to take and use the gratuity after one year, the total reductions at the end of the year are
\begin{align*}
R_1=\frac{n}{r_c}\left(\frac{R}{mn}-\frac{r_c}{n}L\right)\left[\left(1+\frac{r_c}{n}\right)^n-1\right]+[1-\delta(1-q)]G.
\end{align*}
Alternatively, if one opts to take the gratuity by installments and use all of it to repay the loan, the total loan reduction after one year is
\begin{align*}
R_2=\frac{n}{r_c}\left[\frac{R}{mn}+(1-\delta)\frac{G}{n}-\frac{r_c}{n}L\right]\left[\left(1+\frac{r_c}{n}\right)^n-1\right].
\end{align*}
It follows that $R_1-R_2=\phi(r_c)G$, where
\begin{align*}
\phi(r_c)=[1-(1-q)\delta]-\frac{(1-\delta)}{r_c}\left[\left(1+\frac{r_c}{n}\right)^n-1\right].
\end{align*}
The function $\phi(r_c)$ is a {\em decision function} that can be used to determine whether or not to take the gratuity by installments. If $\phi(r_c)<0$, then it is better to take the gratuity by installments. Otherwise it is better to wait for a year before taking the gratuity.
\begin{figure}
\begin{center}
\includegraphics[height=8.5cm,width=8.5cm]{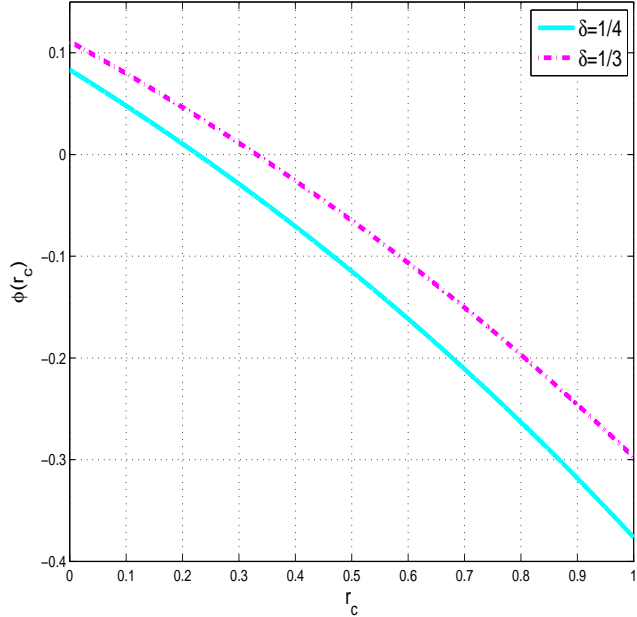}
\caption{\em\small Graphs of the decision function for use in determining whether to help repay a loan by installments of the gratuity or to wait for the end of the year to receive the gratuity as a lump sum to help repay the loan. The graphs correspond to two different tax rates on the gratuity.}
\label{fig:loan}
\end{center}
\end{figure}
Graphs of this decision function are given in Figure~\ref{fig:loan} for parameter values $q=1/3$, $n=12$ and for two values of the tax rate $\delta=1/4,1/3$. For the tax rate of $\delta=1/4$, which is applicable to a wide range of employees in Botswana, $\phi(r_c)>0$ for $r_c<0.225$ and negative otherwise. This implies that for loans with an interest rate less than 22.5\%, it is better to receive the gratuity annually than by monthly installments. On the other other hand, when the loan interest rate exceeds 22.5\%, then it is beneficial to receive the gratuity monthly and use it to help repay the loan. The graph corresponding to $\delta=1/3$ indicates that increasing the tax rate increases the threshold, whilst the qualitative results remain the same.
\section{Discussion}
This article considered the issue of deciding whether to receive gratuity monthly or at the end of the year. In addressing this issue, we first considered the hypothetical and yet very attractive option of one being given all the gratuity at the beginning of the year. To an employee, this is the best offer if one wants to receive the gratuity before the end of the year. Is waiting for one year before receiving the gratuity to get a tax relief on a fraction of it worth the wait? Many could be tempted to jump at this seemingly glorious opportunity to get the gratuity now. After all, an amount of money spent today could be worth more than a greater amount in the future. For specific parameter values applicable to a wide range of employees in Botswana, our analysis showed that if this gratuity was to be invested in a risk-free account, the interest offered should be at least 11.11\% under simple interest. Under continuous compounding, the interest rate offered on the account should be at least 10.54\%. In Botswana, average inflation is about 6\% whilst it is very hard to get a bank interest rate exceeding 5\%. In light of these observations, the seemingly glorious opportunity to receive all the gratuity at the beginning of the year is not economically sound. One can substantially beat inflation by just waiting until the end of the year. Some people might have ideas that have potential to add more value to their gratuity (if received a year earlier) than the minimum required interest rate computed here. To such people, receiving the gratuity at the beginning of the year could be the better option.

In realistic situations, employees are offered the option to receive their gratuity in equal monthly installments rather than as a lump sum at the beginning of the year. This situation was addressed by first tackling the case of an arbitrary number of equal installments in a year. Under this general scenario, we separately considered the cases of simple interest and compound interest. Some general formula was then derived in the simple interest case. Based on the formula obtained, the following points can immediately be noted. Increasing either the tax rate or the fraction that is exempt from tax when gratuity is received at the end of one year increases the interest rate necessary to make taking gratuity by installments attractive. The formula can be used to plug in parameter values applicable to different circumstances. Using parameter values applicable to a range of employees in Botswana, it was found that the minimum interest rate necessary to match receiving gratuity at the end of one year is 20.51\% under simple interest and 19.17\% under continuously compounded interest. These rates are way above what any bank would be willing to give. Moreover, the gain that one might get from counteracting inflation fades into insignificance. The point here is that it is even far better to wait than to receive gratuity by monthly installments. Our findings indicate that taking gratuity by monthly installments is tantamount to losing about 20\% of one's potential net income.

We also considered the situation where one takes a loan at an interest compounded monthly. The decision question was: How should one receive the gratuity to help repay the loan. It was found that in this case, there is a critical loan interest rate (22.5\% for Botswana) below which it is better to wait to receive the gratuity at the end of the year. In Botswana, it is rare for banks to offer loans at annual interest rates above the threshold of 22.5\%. Nonetheless, if the annual interest rate is above the threshold, then taking the gratuity by monthly installments counters the interest rate better than waiting until the end of the year.
\section{Conclusions}
This paper discussed whether employees on fixed-term contracts should opt to receive their employment gratuity by monthly installments or to wait until the end of the year to take it as a lump sum. This problem was considered in two contexts: 1) depositing the installments into a risk-free savings account and 2) using the installments to help repay a loan. In each case, taking the gratuity by monthly installments can at least match the benefit of end of year tax relief only if the interest rate exceeds a certain threshold. For a wide range of employees in Botswana, the necessary threshold is well above the interest rates that banks typically give or charge. The results can be applied beyond saving the installments in a risk-free savings account or helping repay a loan. Other people may be interested to use the installments in other investments like renovating a house. The point of this paper is that using the installments in such an investment is economically sound if it results in returns that exceed the threshold. In the case of Botswana, the results suggest that (for a wide range of employees) taking the gratuity by monthly installments could be tantamount to loss of income.

Whilst we have drawn parallels between employment gratuity and pension schemes, we stress that the eagerness for individuals to get their dues by monthly installments does not imply that they would not be interested in an undisguised pension scheme. Some might even take a portion of their monthly gratuity into a pension plan. Furthermore, it might be possible for employees to be persuaded by the findings of this paper to opt for a yearly plan instead of a monthly one. The persuasion of this paper stands on the platform of the certainty of employment gratuity, a feature that pension schemes lack.
\section*{Acknowledgements}
The author would like to acknowledge fruitful discussions with G. Malumbela, E. Masuku and M. McGeoff, colleagues at BITRI in Botswana.
\bibliographystyle{pef}
\bibliography{refs}
\end{document}